\documentclass[12pt,tightenlines,eqsecnum,floats,shownopacs,nofootinbib,amsmath,amssymb,aps,prd]{revtex4}

\usepackage{color}
\usepackage{graphicx}
\usepackage{amsmath,amssymb}

\def\be{\nopagebreak[3]\begin{equation}}
\def\ee{\end{equation}}
\def\ba{\nopagebreak[3]\begin{eqnarray}}
\def\ea{\end{eqnarray}}

\def\f{\frac}

\def\h{\hat}

\def\s{\sigma}
\def\S{\mathcal{S}}

\def\T{{T}}

\def\H{\mathcal{H}}

\def\U{\mathcal{U}}
\def\O{\mathcal{O}}
\def\Tr{\rm Tr}

\begin{document}

\title{A new perspective on $CP$ and $\T$ violation}
\author{Abhay Ashtekar}
\email{ashtekar@gravity.psu.edu} \affiliation{Institute for
Gravitation and the Cosmos \& Physics Department, Penn State,
University Park, PA 16802, U.S.A.}

\begin{abstract}

It is shown that the results of $CP$ and $\T$ violation experiments
can be interpreted using a very general framework that does not
require a Hilbert space of states or linear operators to represent
the symmetries or dynamics. Analysis within this general framework
brings out the aspects of quantum mechanics that are essential for
this interpretation. More importantly, should quantum mechanics be
eventually replaced by a new paradigm, this framework could
\emph{still} be used to establish violation of $CP$ and $\T$
invariance from the already known experimental results.

This work arose as a `formal response' to a talk \emph{Three Merry
Roads to T-Violation} by Dr. Bryan Roberts [1]. Both talks were
given at a \emph{Workshop on Cosmology and Time} held at Penn State
in April 2013, which brought together physicists and philosophers of
science.

\end{abstract}

\maketitle

\section{Introduction}
\label{s1}

To interpret any experiment, one needs a theoretical paradigm. For
experiments on violations of the $CP$ and  $\T$ symmetries in weak
interactions, the standard choice has been the mathematical
framework of quantum physics. This is natural because the best
available description of the electroweak interactions is based on
local quantum field theory. However, we will show that one can in
fact analyze the $\T$-violation experiments in a \emph{much} more
general setting. Classical mechanics, quantum mechanics and quantum
field theory provide only specific examples of this general
framework. As is often the case, because this framework has much
less structure, the analysis becomes significantly simpler. As a
result, aspects of quantum physics that are truly essential for
interpreting experiments on $CP$ and $\T$-violation are brought to
the forefront. They are clearly separated from other features which,
though not central, have generally been treated in the literature as
being equally significant. More importantly, the conceptual
structure that is essential to the interpretation of experiments is
so weak that, even if quantum physics were to be replaced by some
more general framework ---e.g., because of its unification with
general relativity--- the current experiments will still, in all
likelihood, enable us to conclude that the fundamental laws of
Nature fail to be invariant under time reversal.

In the tradition of joint physics and philosophy meetings, this is a
formal `response' to a paper \cite{bwr} presented by Dr. Roberts in
which he crystalized the conceptual underpinnings of experiments
that established the $CP$ and $\T$ violations. His clear
presentation led me to sharpen my own understanding of the
theoretical constructs that have been used to analyze these
experiments. This communication is the result of that
re-examination. In Dr. Roberts'  terminology, the analysis of the
Cronin-Fitch experiment on $CP$-violation can be based on \emph{the
Curie principle} and that of the CPLEAR experiments on
$T$-violation, on \emph{the Kabir Principle}. Therefore I will
discuss these principles in detail in sections \ref{s2} and
\ref{s3}. I have attempted to make this discussion reasonably
self-contained but, to fully appreciate its content, it would be
helpful if the reader has already gone through Dr. Robert's paper
\cite{bwr}. (There is also interesting  follow-up paper \cite{bwr2}
in which Dr. Roberts has used the framework introduced in the next
two sections to generalize a third approach to $\T$-violation.)

The issue of time-reversal has two distinct facets: microscopic and
macroscopic. The microscopic $\T$-violation discussed in \cite{bwr}
and in the present communication is quite distinct from the manifest
arrow of time we perceive in our everyday life and, more generally,
in the physics of large systems. Since Dr. Roberts mentioned the
macro-world only in passing, before entering the main discussion,
let me begin with a brief detour to explain this point. For
simplicity, I will use the framework of classical physics because
the core of the argument is not sensitive to the distinction between
classical and quantum mechanics. Consider a large box with a
partition that divides it into two parts, say, the right and the
left halves. Suppose there is some gas in the left half and vacuum
in the right. Once the equilibrium is reached, the macroscopic state
of this gas is described by the volume it occupies, $V_i$; the
pressure it exerts on the walls of the box, $P_i$; and its
temperature $T_i$; where $i$ stands for `initial'. If we open the
partition slowly, the gas will fill the whole box and its
macro-state in equilibrium will be described by new parameters,
$V_f, P_f, T_f$. Thus, there has been a transition from the initial
macro-state $(V_i, P_i, T_i)$ to a final state $(V_f, P_f, T_f)$.
Our common experience tells us that the time reverse of this process
is \emph{extremely} unlikely. However, we also know that the
microscopic variables for the system are the positions and momenta
of some $10^{23}$ molecules in the box. These are subject just to
Newton's laws which are \emph{manifestly invariant under the
time-reversal operation} $\T$! Therefore, if we were to reverse the
momentum $\vec{p}{}_{(\alpha)}(t)$ of each molecule (labeled by
$\alpha$) at a late time $t$, keeping the positions
$\vec{x}{}_{(\alpha)}(t)$ the same, time evolution would indeed
shift the gas back from its final macroscopic state to the initial
one, confined just to the left half of the box. But in practice it
is very difficult to construct this time-reversed initial state.
Thus, there is indeed a macroscopic arrow of time but its origin is
\emph{not} in the failure of the microscopic laws to be invariant
under $\T$ but rather in the fact that the initial conditions we
normally encounter are very special. Specifically, in our example,
there are vastly fewer micro-states compatible with the initial
macro-state $(V_i, P_i, T_i)$ than those
that are compatible with the final macro-state $(V_f, P_f, T_f)$.%
\footnote{This is primarily because the volume $V_f$ allowed for
\emph{each} molecule in the final macro-state is twice as large as
$V_i$, allowed in the initial macro-state. Consequently, the number
of microscopic configurations compatible with the final macroscopic
state is about $2^{10^{23}}$ times the number of microscopic states
compatible with the initial macroscopic configuration.}
Put differently, the entropy of the initial macro-state is much
lower than that in the final macro-state. This is the arrow of time,
commonly discussed in the literature.

It is clear from the simple example that the fact that there is an
obvious arrow of time in the macro-world does \emph{not} imply that
the microscopic or fundamental laws have to break $\T$-invariance.
Indeed, as Dr. Roberts emphasized in the beginning of his article
\cite{bwr}, it was common to assume that the fundamental laws
\emph{are} invariant under the time-reversal operation $\T$. It came
as a shock that the weak force violates this `self-evident' premise!

The rest of this article will focus on the $\T$ violation at the
fundamental, microscopic level.

\section{Weak Interactions and the Curie Principle}
\label{s2}

As Dr. Roberts has explained, what the Cronin-Fitch experiment
\cite{cf} establishes directly is that the weak interactions are not
invariant under $CP$, i.e., under the simultaneous operations of
charge conjugation $C$ and reflection through a mirror, $P$.  As
normally formulated, the parity operation is meaningful only if the
underlying space-time is flat, i.e., represented by Minkowski
space-time. This means one ignores curvature and therefore gravity.
One further assumes that physics is described by a local quantum
field theory on this Minkowski space, for which individual physical
fields transform covariantly under the action of the Lorentz group,
and dynamics is generated by a self-adjoint Hamiltonian obtained by
integrating a scalar density (or a 3-form), constructed locally from
the physical fields. Then, one has the \emph{CPT theorem} that
guarantees that the product $CPT$ of charge conjugation, $C$,
parity, $P$ and time reversal $\T$ is an exact \emph{dynamical} symmetry.%
\footnote{This is a summary of the discussion one finds in quantum
field theory text books (see, e.g. \cite{sw}). More rigorous
versions based on Weightman axioms \cite{s-w} and the algebraic
approach \cite{yb} are also available in the literature. However,
the self-adjoint Hamiltonian required in the text book version is
not rigorously defined in four dimensions beyond the quantum theory
of free fields. Similarly, we still do not have a single example of
a 4-dimensional, interacting quantum field theory satisfying either
the Wightman axioms or the axioms of the algebraic quantum field
theory. Thus, there is a curious mis-match between the mathematical
statements of CPT theorems and theories of direct physical
interest.}
Therefore, although the Cronin-Fitch experiment does not
\emph{directly} imply $\T$-violation, as Dr. Roberts explained, if
we assume that weak interactions are described by a local quantum
field theory in Minkowski space, then the observed breakdown of CP
invariance implies that they violate $\T$ invariance as well. In the
rest of this section, I will focus on just the $CP$ symmetry and its
violation observed in the Croin-Fitch experiment. Thus, I will not
need to refer to the CPT at all.

In the current analysis of $CP$ violation, one uses the following
form of the Curie principle: If an initial state $\s_i$ is in
variant under $CP$ but its time-evolved final state $\s_f$ is not,
then dynamics can not be $CP$ invariant. As explained in section 2.4
of \cite{bwr}, the analysis has the remarkable feature that it does
not assume a specific Hamiltonian $H$. Therefore, the argument will
remain unchanged should we discover that the currently used
Hamiltonian $H$ in electro-weak interactions has to be modified,
e.g., to accommodate future experiments, or to unify them with
strong interactions.

However, the standard analysis \emph{does} make a crucial use of the
detailed kinematical structure of quantum physics (summarized
below). If a future quantum gravity theory were to require that this
structure has to be modified ---e.g., by removing the emphasis on
linearity--- then the standard analysis cannot be used to conclude
that the Cronin-Fitch experiment implies a violation of $CP$
invariance in weak interactions. \emph{The main point of this
section is to show that this specific kinematical framework of
Hilbert spaces and operators is not really necessary.} The Curie
principle can be extended to a much more general framework than that
offered by quantum physics.

Let us begin by introducing this framework, which we will call
\emph{general mechanics}. It will incorporate quantum as well as
classical mechanics, but only as specific special cases. The basic
assumptions of \emph{general mechanics} are:

\begin{itemize}
\item i) We have a set $\S$ of states;
\item ii) There is a 1-1, onto dynamical map $S$ ---the
    `$S$-matrix'--- from $\S$ to itself. This $S$ could refer to
    finite time evolution, say from time $t_1$ to $t_2$ or,
    alternatively, to the time evolution in the infinite past to
    the infinite future. In practice is it convenient to
    consider two copies $\S_{i}$ and $\S_{f}$ of $\S$,
    representing initial and final states, and regard $S$ as a
    map from $\S_{i}$ to $\S_{f}$;
\be S: \S_i \rightarrow \S_f; \quad\quad S(\s_i) = \s_f
\quad\quad \forall \s_i \in \S_i \ee
\item iii) Potential symmetries are represented by a 1-1, onto
    maps $R: \S \to \S$, from $\S$ to itself. We will first
    consider the case in which $R$ maps $\S_i$ to itself and
    $\S_f$ to itself. This is the case if $R$ is, for example,
    the discrete symmetry represented by $C$, or $P$ or $CP$.

\end{itemize}

Note that framework is truly minimalistic. \emph{The space $\S$ of
states is just a set}; no further structure is needed. To draw the
contrast, let me summarize the structure we use
in classical and quantum mechanics.%
\footnote{But the material till the end of this paragraph can be
skipped without loss of continuity. It brings to forefront the
intricacy of the structure underlying classical and quantum
mechanics, in contrast to \emph{general mechanics}. }
In classical mechanics, $\S$ is the phase space which is assumed to
be a smooth, even dimensional differentiable manifold $\Gamma$,
equipped with a non-degenerate, closed 2-form $\Omega$, called the
symplectic structure. In quantum mechanics, one first introduces an
auxiliary structure, a Hilbert space, $H$: it is a complex vector
space that is endowed with an Hermitian scalar product and is
complete in the sense that the limit point of each Cauchy sequence
in $H$ is again in $H$. Pure states are then represented by rays in
$H$ and mixed states by density matrices, i.e., bounded,
self-adjoint operators with finite trace. As for dynamics, in
classical mechanics, it is given by the flow generated by a
Hamiltonian vector field $X = \Omega^{-1}\, \urcorner\, dH$ for a
smooth function $H$ on $\Gamma$, called the Hamiltonian. Now the map
$S$ corresponds either to the finite $S$-matrix between a time $t_1$
and $t_2$, obtained by moving along the Hamiltonian flow an affine
parameter distance $t_2 - t_1$, or its limit as $t_1 \to -\infty$
and $t_2 \to \infty$. In quantum mechanics, as mentioned above,
dynamics is represented by a unitary flow $\U_t = \exp -itH$
generated by a self-adjoint operator. The map $S$  corresponds again
either to a finite $S$ matrix,\, $\U_{t_2-t_1}$,\, or its limit as
$t_1 \to -\infty$ and $t_2 \to \infty$. Finally, in classical
mechanics the potential symmetries $R$ are diffeomorphisms on
$\Gamma$ that preserve the symplectic structure $\Omega$ and those
in quantum mechanics are linear, unitary mappings on $H$.

This rather detailed discussion serves to brings out three points.
First, both classical and quantum mechanics are based on very rich
but specific and intricate mathematical frameworks. From the summary
of the literature contained in section 2.5 of \cite{bwr}, it would
appear that the specifics of quantum mechanics ---the Hilbert space
structure of $\H$ and linearity of the candidate symmetries $R$---
are all essential for a formulation of the Curie principle that can
be used to interpret the Cronin-Fitch experiment. Second, the
mathematical underpinnings of the classical and the quantum
frameworks are \emph{very different from each other.} Therefore, it
is natural to anticipate that the next paradigm shift (that may
emerge, e.g., from a successful quantum gravity theory) could bring
with it another rich yet \emph{different} mathematical framework.
Would one then have to reconstruct an appropriate formulation of the
Curie principle and re-examine experiments using the new framework?
The third point is that this would not be necessary if one could
develop a formulation of the Curie principle (and also the Kabir
principle, discussed in section \ref{s3}) using just the three
assumptions of \emph{general mechanics} which have survived the
transition from classical physics to quantum. Indeed, the
assumptions are so minimalistic that it is reasonable to hope that
they will also be realized in the future, more general mechanics.

Let us then investigate if the Curie principle can be formulated
using just the three assumptions of \emph{general mechanics}. In the
spirit of this principle, suppose that there exists some initial
state $\s_i \in \S$ such that
\be R \s_i = \s_i \quad \quad {\rm but}\quad \quad R \s_f \not=
\s_f\ee
Then,
\be R\, (S \s_i) = R\s_f \not= \s_f, \quad {\rm but}\quad \s_f =
S(\s_i) = S\, (R\s_i) \, . \ee
Therefore, we conclude
\be  SR \not= RS \, . \ee
Thus, the dynamical map $S$ does not commute with the candidate
symmetry $R$: \, $R$ \emph{can not a dynamical symmetry}. To
summarize, we have derived the Kabir principle in the framework of
\emph{general mechanics}: \emph{If there exists a state $\s_i$ such
that $R \s_i = \s_i$\, and\, $S \s_i = \s_f$ but $R \s_f \not= \s_f$
then $R$ is not a dynamical symmetry of the system.}%
\footnote{It is straightforward to alter the argument to obtain the
other desired conclusion of fact 1 in section 2.4 of \cite{bwr}.}
By repeating the argument of \cite{bwr}, now it is easy to analyze
the Cronin-Fitch experiment in the setting of \emph{general
mechanics}. Even if the theory describing weak interaction is more
general than quantum mechanics, the experiment would still imply
that the dynamics of weak interactions violates $\T$-invariance, so
long as the underlying framework of this theory meets the three
basic postulates of \emph{general mechanics}.

Note that this general formulation in particular enables one to
discuss in one go all the symmetries $R$ in classical mechanics and
linear symmetries used in quantum mechanics. Furthermore, since one
does not even need to refer to the underlying Hilbert space
structure, one can also apply the argument within standard quantum
mechanics even if $R$ or $S$ were represented by \emph{non-linear}
mappings. In particular, for the argument to hold, the $CP$ symmetry
does not have to be unitarily realized. By contrast, as summarized
in \cite{bwr}, the current literature on the Cronin-Fitch experiment
often emphasizes that the assumption of unitarity is essential to
conclude that the experiment implies $CP$ violation.

We will conclude this discussion of the Curie principle with three
remarks.

1) Since $\S$ is only a set, we cannot speak of continuous evolution
in the framework introduced above. But one could trivially extend
the framework by endowing $\S$ with a topology and replacing $S$
with a continuous evolution map $E(t)$, where $t$ is to be thought
of as a time parameter. The argument given above will then imply
that $E(t)$ will not commute with $R$.

2) However, we \emph{did} assume that $R$ maps the space $\S_i$ of
initial states to itself and the space $\S_f$ of final states to
itself. This assumption is \emph{not} satisfied by the time-reversal
operation $\T$, which maps initial states to final states (and vice
versa): $\T: \S_i \to \S_f$ is a 1-1 onto map from $S_i$ to $S_f$.
Therefore, in this case, $\T$ is a dynamical symmetry if and only if
\be S\s_i = \s_f\quad \implies  \quad  S^{-1} (\T\s_i) = \T^{-1}
(\s_f) =  \T^{-1} (S\s_i) \ee
i.e., the time reverse of $\s_i$ (which is in $\S_f$) is mapped by
dynamics to the time reverse of $\s_f$. The generalization of the
Curie principle discussed above does not have any implication in
this case. Therefore to directly test $\T$-invariance we need a
generalization of the Kabir principle, discussed in section
\ref{s3}. In this respect, the situation is the same as in section 2
of \cite{bwr}.

3) While $R$ invariance of dynamics is captured by the property $RS
= SR$ of the S-matrix $S$, the $\T$ invariance is captured by
$S^{-1}\T = \T^{-1} S$. As we just saw, the difference arises simply
because, while $R$ preserves each of $\S_i$ and $\S_f$,\, $\T$ maps
one to the other. At a fundamental level, then, \emph{the difference
is not because $R$ is linear while $\T$ is anti-linear,} as is often
stated in the literature: If one works entirely in the quantum
framework, one misses the broader perspective and is then tempted to
focus only on the anti-linearity of $\T$.

\section{The Kabir Principle}
\label{s3}

By using the Curie principle in conjunction with the Cronin-Fitch
experiment we have concluded that $CP$ cannot be a dynamical
symmetry in any theory of weak interactions that conforms to the
principles of \emph{general mechanics}. But this provides only an
indirect evidence for $\T$-violation because one needs to assume, in
addition, validity of the $CPT$ theorem. As explained in section 3
of \cite{bwr}, more direct evidence for $\T$-violation comes from
another idea, called the Kabir principle by Roberts, which is again
formulated in the framework of quantum mechanics: If a transition
$\s_i \to \s_f$ occurs with a different probability from the
transition $\T \s_f \to T \s_i$ then the dynamics underlying this
process cannot be invariant under the time-reversal operation, $\T$.
Again this argument is deeply rooted in the detailed structure of
quantum mechanics. A natural question for us is then: Can we extend
the Kabir principle to \emph{general mechanics}? The answer is in
the affirmative. However, since it refers to probabilities, even to
state Kabir's principle one needs to introduce some additional
structure in \emph{general mechanics}.

Let us then introduce the necessary structure, in addition to the
three postulates stated in section \ref{s2}:
\begin{itemize}
\item An \emph{overlap map} $\O$ on the space of states $\S$
    (and therefore on each of $\S_i$ and $\S_f$):\,\, $\O : \S
    \times \S \rightarrow [0,1] \in R$, such that
\be \label{sym} \O (\s,\, \s^\prime) = \O (\s^\prime,
\s),\quad\quad \forall\,\, \s, \s^\prime \in \S. \ee
$\O (\s, \s^\prime)$ is to be thought of as the overlap between
states $\s$ and $\s^\prime$.
\end{itemize}
In classical mechanics, general states are represented by
non-negative distributions $\rho$ on the phase space $\Gamma$ which
are normalized so that $\int_\Gamma \rho\, dV_\Omega = 1$, where
$dV_\Omega$ is the Liouville volume element on $\Gamma$. Then the
overlap function $\O(\rho, \rho^\prime)$ can be the obvious overlap
between the two states: $\O(\rho, \rho^\prime) = \int_\Gamma (\rho
\rho^\prime)^{\f{1}{2}} \, dV_\Omega$.%
\footnote{The square-root in the integrand ensures that the overlap
function is bounded by $1$. (It is not needed in quantum theory.) A
pure state $\rho = \rho_\gamma$, is just the Dirac distribution
peaked at a point $\gamma$ of $\Gamma$. For these states, natural
regularizations yield : $\O(\rho_{\gamma}, \rho_{\gamma^\prime}) =
\delta_{\gamma,\, \gamma^\prime}$. Thus the overlap between a pure
state and a distinct pure state simply vanishes while its overlap
with itself is perfect.}
In quantum mechanics, a general state is represented by a normalized
density matrix $\h\rho$ ---a non-negative, bounded, self-adjoint
operator on the Hilbert space $\H$ with $\Tr \h\rho =1$,\, where
$\Tr$ is the trace operation. The overlap function can be then taken
to be $\O (\h\rho,\, \h{\rho}^\prime ) = \Tr \,
\h\rho\,\h{\rho}^\prime$. Pure states
---projection operators on one dimensional subspaces of $\H$--- can
be represented by (equivalence classes of) normalized elements
$\psi$ of $\H$ (where two elements are equivalent if they differ by
a phase). Then the overlap function $\O(\psi,\, \psi^\prime)$ is
simply the standard overlap probability of quantum mechanics:
$\O(\psi,\, \psi^\prime) = |\langle \psi, \psi^\prime \rangle|^2$.

Let us return to \emph{general mechanics.} Since the overlap map is
part of the kinematical structure, the notion of symmetries and
dynamics, introduced in section \ref{s2}, have to be compatible with
it. The compatibility conditions are all natural and obvious. But
since they will be used in the proof of the Kabir principle, let us
spell them out explicitly.

\begin{itemize}
\item To qualify as symmetry, the map $R:\, \S_i\to \S_i $
    introduced in section \ref{s2} has to satisfy
\be \O_i(R\s_i,\, R\s^\prime_i) = \O_i (\s_i,\, \s^\prime_i)\ee
on $\S_i$ (and similarly on $\S_f$). Similarly to qualify as
symmetry, the time reversal map $\T:\, \S_i \to \S_f$ must
satisfy
\be \label{T} \O_f (\T\s_i,\, \T\s^\prime_i) = \O_i (\s_i,\,
\s^\prime_i)\ee
for all $\s_i \in \S_i$.

\item The dynamical map $S: \S_i \rightarrow \S_f$ should  be
    compatible with the overlap map, i.e., satisfy
\be \label{S}\O_i (\s_i,\, \s^\prime_i) = \O_f (S\s_i,\,
S\s^\prime_i) \equiv \O_f (\s_f,\, \s^\prime_f)\, . \ee
\end{itemize}

Finally, using the overlap map $\O$ and the dynamical map $S$, we
introduce the notion of transition probability:
\begin{itemize}

\item Given a dynamical map $S$, the \emph{transition
    probability} between an initial state $\s_i \in \S_i$ and
    \emph{any} given final state $\s_f^\prime$ is defined to be
\be \label{prob} P(\s_f^\prime, \, \s_i) := \O_f (\s_f^\prime,
\, S\s_i)\, \equiv \,\O_f(\s_f^\prime, \, \s_f) \, . \ee
\end{itemize}

This concludes the additional kinematical structure we need on our
\emph{general mechanics} to formulate the Kabir principle. Note that
the notions of the overlap function and transition probability are
already available in classical as well as quantum mechanics. In both
cases symmetries and dynamics satisfy the compatibility conditions
we have introduced. Thus, as in section \ref{s2}, \emph{general
mechanics} continues to be minimalist: it simply extracts certain
notions that are common to both classical and quantum paradigms,
which we can reasonably expect to be present also in the future
generalizations of these settings.\\

Recall that $\T$ is a dynamical symmetry if and only if $S^{-1}\T =
\T^{-1}S$, or, equivalently,  $ST^{-1} = TS^{-1}$. \emph{Suppose a
dynamical map $S$ satisfies this condition.} As before, let us use
the notation $\s_f = S\s_i$ and $S\s_i^\prime = S\s_f^\prime$. Then,
\be \O_f (T\s_i,\, S (T^{-1}\s_f^\prime)) = \O_f (T\s_i,\, T
(S^{-1}\s_f^\prime)) = \O_f (T\s_i,\, T\s_i^\prime) = \O_i
(\s_i,\, \s_i^\prime) \ee
where in the last step we used (\ref{T}). On the other hand,
(\ref{S}) and (\ref{sym}) imply
\be \O_i (\s_i,\, \s_i^\prime) = \O_f (S\s_i,\, S\s_i^\prime) =
\O_f(S\s_i^\prime, S\s_i) = \O_f (\s_f^\prime, S \s_i). \ee
The last two equations and the definition (\ref{prob}) of transition
probability now yield
\be P(\s_f^\prime, \s_i ) = P(T\s_i,\, T^{-1}\s_f^\prime)\, .\ee

Thus, we have shown that if $\T$ is a symmetry of the dynamical map
$S$ then the transition probability between the states $\s_i$ and
$\s_f^\prime$ must equal that between the two states
$T^{-1}\s_f^\prime$ and $T\s_i$, obtained by a time-reversal.
Therefore if the transition probability between \emph{any} two
states and their time reversed versions differ observationally, then
the time reversal
symmetry is broken by dynamics.%
\footnote{It is worth noting that, in practice, one measures is the
transition rate. This is not determined solely by the transition
probability. In the leading order approximation (``Fermi's golden
rule'') the transition probability has to be multiplied by the
density of final states. But in practice one can easily take care of
this issue and verify whether or not the transition probability is
symmetric under time-reversal.}
This is the generalization of the Kabir principle from quantum
mechanics to \emph{general mechanics} we were seeking. As Dr.
Roberts has explained, it has been used to show that the \emph{kaon
oscillations}, $K^0\, \leftrightarrows\, \bar{K}^0$ observed at the
CPLEAR detector provide a direct violation of $\T$-invariance ( see,
e.g., \cite{cplear,alvarez,ellis}). Using the same arguments as in
section 3.2 of \cite{bwr}, we can now conclude that the CPLEAR
observations imply that the time-reversal symmetry $\T$ is violated
in \emph{any} theory of weak interactions \emph{so long as it falls
in the broad framework of general mechanics} developed in
this section.\\

As with the discussion of the generalized Curie principle of section
\ref{s2}, this generalization of the Kabir criterion does not refer
to the Hilbert space structure of the space of states or linearity
(or anti-linearity) of various maps. In particular, in the case of
quantum mechanics, while it incorporates the standard treatment
neatly summarized in section 3 of \cite{bwr}, the result would hold
even if, say, the S-matrix were anti-unitary (or, indeed,
non-linear!). By contrast, as emphasized in \cite{bwr}, it is
generally believed that unitarity of the $S$ matrix is essential to
conclude that we have experimental evidence of direct $\T$-violation
in weak interactions.

\section{Discussion}
\label{s4}

Already in the context of quantum mechanics, it is interesting to
analyze the interplay between symmetries and dynamics using only
that structure which is available in \emph{general mechanics}. The
analysis becomes much simpler and, furthermore, brings to forefront
the concepts and structures that are essential in the discussions of
$CP$ and $\T$ violations, separating them from inessential notions
that have added unnecessary complexity to the analyses one finds in
the literature. In particular, neither the linear structure nor the
details of the Hermitian inner product on the space of quantum
mechanical states is essential. Nor is it necessary that symmetries
have to be represented by unitary or anti-unitary operators; indeed,
they need not even be linear operators! Secondly, the primary
distinction between $\T$ on the one hand, and $C$, $P$, or $CP$ on
the other, is only that while the spaces of `in' and `out' states
are separately invariant under $C$, $P$ and $CP$, the map $\T$ sends
the incoming states to the outgoing ones. In standard quantum
mechanics, $C$, $P$ and $CP$ are represented by linear, unitary
maps, while $\T$ is represented by an anti-linear, anti-unitary map.
However, if one uses only that structure which is provided by
\emph{general mechanics} one finds that, contrary to a wide-spread
belief, this difference is not primary to the distinction between
the Curie and the Kabir criteria.

While these points are illuminating, the main strength of the more
general perspective presented here is that it broadens the reach of
the Cronin-Fitch and CPLEAR experiments enormously. We are no longer
constrained to analyze them in the strict framework of quantum
mechanics. This is a significant advantage because future advances,
particularly in quantum gravity, may well lead us to a mathematical
framework which is quite different from that offered by quantum
mechanics.  Just as in the passage from special relativity to
general relativity we had to abandon flatness of space-time
geometry, the future theory may well force us to abandon the linear
structure that underlies quantum mechanics. For example, in quantum
mechanics the pure states are represented by rays in the Hilbert
space $\H$ and constitute a trivial K\"ahler manifold. It may be
replaced by \emph{general, non-trivial} K\"ahler manifolds which do
\emph{not} refer to a Hilbert space. Or, dynamics and symmetries may
be represented by \emph{non-linear} maps. Indeed, there are concrete
ideas as to how this may come about (see, e.g., \cite{as} and
references therein). But such generalizations are likely to retain
the minimal structure of \emph{general mechanics}, introduced in
sections \ref{s2} and \ref{s3}. If so, then the current experiments
will still continue to imply that the dynamics of weak interactions
violates $CP$ and $\T$ invariance even in these much more general
theories. Thus, at a conceptual level the reach of the current
experiments on $CP$ and $\T$ violations seems to be extraordinarily
long, going well beyond the standard quantum mechanics framework in
which they have generally been interpreted.

\section*{Acknowledgments}

I would like to thank Bryan Roberts for stimulating discussions,
John Collins for a valuable e-mail exchange and Emily Grosholz, John
Norton and Tom Pashby for raising insightful questions that have
improved this presentation. Additional thanks are due to Emily
Grosholz for her leadership in bringing together physicists and
philosophers for a lively workshop. This work was supported in part
by the NSF grant PHY-1205388 and the Eberly research funds of Penn
state.

\vfill\break

\end{document}